\documentclass[a4paper, preprint, jcp]{revtex4}
\usepackage{amsmath}
\usepackage{amsfonts}
\usepackage{graphicx}
\usepackage{color}
\usepackage{lscape}
\usepackage{amssymb}

\begin{document}

\title{Dynamical simulations of polaron transport in conjugated polymers with the inclusion of electron-electron interactions}
\author{Haibo Ma}
\email{haiboma@physik.rwth-aachen.de}
\author{Ulrich Schollw\"{o}ck}
\email{scholl@physik.rwth-aachen.de} \affiliation{Institute for Theoretical Physics C, RWTH Aachen University, D-52056 Aachen,
Germany}
\date{Latest revised on \today}

\begin{abstract}

Dynamical simulations of polaron transport in conjugated polymers in the presence of an external time-dependent electric field have been performed within a combined extended Hubbard model (EHM) and Su-Schrieffer-Heeger (SSH) model.
Nearly all relevant electron-phonon and electron-electron interactions are
fully taken into account by solving the time-dependent Schr\"{o}dinger equation for the $\pi$-electrons and the Newton's equation of motion for the backbone monomer displacements by virtue of the combination of the adaptive
time-dependent density matrix renormalization group (TDDMRG) and classical molecular dynamics (MD).
We find that after a smooth turn-on of the external electric field the polaron is accelerated at
first and then moves with a nearly constant velocity as one entity consisting of both the charge and the lattice deformation. An ohmic region (3
mV/$\text{\AA}$ $\leq E_0\leq$ 9 mV/$\text{\AA}$) where the
stationary velocity increases linearly with the electric field
strength is observed for the case of $U$=2.0 eV and $V$=1.0 eV. The maximal velocity is well above the speed of sound. Below 3 mV/$\text{\AA}$ the polaron velocity increases nonlinearly and in high electric fields with strength $E_0\geq$ 10.0 mV/$\text{\AA}$ the polaron will become unstable and dissociate. 
The relationship between electron-electron
interaction strengths and polaron transport is also studied in
detail. We find that the the on-site Coulomb interactions $U$ will suppress the polaron transport and small nearest-neighbor interactions $V$ values are also not beneficial to the polaronic motion while large $V$ values favor the polaron transport. 

\end{abstract}

\maketitle
\section{Introduction}
Charge transport properties in conjugated polymers have attracted sustained attention from both academic and industrial researchers, since it was discovered in the 1970s that the electrical conductivity
of trans-polyacetylene (PA) can be improved significantly
by doping with strong electron acceptors or donors
\cite{Chiang77, Shirakawa77, Chiang78}. Many electronic devices have been fabricated based on these conjugated polymers.\cite{MacDiarmid97, Burroughes98, Heeger01, Heeger01_2}

All the conducting polymers can generally be sorted into two classes. The first class is trans-polyacetylene, in which there
exists a twofold degeneracy of ground state energy distinguished by the positions of the double and single bonds. The degenerate ground state of trans-polyacetylene leads to solitons \cite{Heeger88} as the important excitations and the dominant charge storage species, which take the form of domain walls separating
different districts of opposite single and double bonds alternation
patterns. The second class of conducting polymers is given by all the other systems except trans-polyacetylene, in which the ground state degeneracy is weakly lifted. So polarons and confined soliton pairs (bipolarons) \cite{Heeger88} instead of solitons are the important excitations and the dominant charge storage configurations for this class of conducting polymers. Nowadays it has been widely accepted
that these quasiparticles (solitons, polarons and bipolarons) are the fundamental charge carriers in conducting polymers. The studies of the dynamics of these charge carriers are therefore of great interest for the purpose of modulating or devising new organic electronic materials based on conjugated polymers.

So far, there have been many extensive theoretical studies
on polaron dynamics in conjugated polymers \cite{Rakhmanova99, Rakhmanova00, Basko02, Johansson01, Johansson04, Arikabe96, Streitwolf98, Yan04, Yu04, Fu06, Liu06, Li06}  based on the Su-Schrieffer-Heeger (SSH) model
\cite{Su79, Su80}, an improved H\"{u}ckel molecular orbital model,
in which $\pi$-electrons are coupled to distortions in the polymer
backbone by the electron-phonon interaction. Rakhmanova and Conwell \cite{Rakhmanova99, Rakhmanova00} have considered the motion of a polaron under low and high electric fields respectively by using an adiabatic simulation in which the electronic energy is treated within the Born-Oppenheimer approximation. Their results show that the polaron can form and move in its entity in low electric fields but dissociates in high fields. The highest electric field strength under which the polaron can sustain is about 6 mV/$\text{\AA}$. Basko and Conwell \cite{Basko02} have also analyzed the dynamics of a polaron and found that the speed of a steadily moving polaron cannot exceed the sound speed ($V_S$). These conclusions are consistent with those from the numerical results presented by Arikabe \textit{et al} \cite{Arikabe96} but inconsistent with those given by Johansson and Stafstr\"{o}m \cite{Johansson01, Johansson04}, who performed a nonadiabatic simulation in which transitions between instantaneous eigenstates are allowed. In Johansson and Stafstr\"{o}m's work, the maximum velocity (about 4$V_S$) is reached at a high electric field strength ($\sim$ 3.5 mV/$\text{\AA}$) and at even higher electric field strengths the polaron will become unstable and dissociate.

In the SSH model only the electron-lattice interactions are considered and the electron-electron interactions are totally ignored. In order to improve this situation, Di \textit{et al} have considered extended Hubbard model (EHM) combined with the SSH model and investigated the dynamics of polarons within this model at the level of an unrestricted Hartree-Fock (UHF) approximation.\cite{Di07} They found that the localization of the polaron is enhanced and the stationary velocity of the polaron is decreased by both the on-site repulsions $U$ and the nearest-neighbor interactions $V$. Furthermore, they found that the local extremum of the stationary velocity of the polaron appears at $U\approx 2V$. However, in recent years a lot of theoretical calculations of the static properties of
conjugated polymers have shown that the electron correlation effect
plays a very important role in determining the behavior of the
charge carriers in conjugated polymers.\cite{Yonemitsu88, Sim91,
Suhai92, Villar92, Rodriguez-Monge95, Hirata95, Bally92, Fulscher95,
Guo97, Perpete99, Fonseca01, Oliveira03, Champagne04, Monev05, Ma05,
Ma06, Hu07} For example, electron correlation is essential in
obtaining correct pictures of the spin polarizations in
trans-polyacetylene.\cite{Ma05} Therefore, theoretical simulations for
polaron dynamics with electron correlations considered, which step
beyond the SCF level, are highly desirable. However, the calculations for
large polymer systems with traditional advanced electron-correlation
methods such as configuration interaction method (CI), multi-configuration self-consistent field method (MCSCF), many-body Moller-Plesset perturbation theory (MPn) and coupled cluster method (CC) are currently still not feasible due to the huge computational
costs. Fortunately, the adaptive time-dependent density-matrix renormalization group (TDDMRG)
method \cite{White04, Daley04, Schollwock05} can be used instead. In the context of 1D correlated electronic and bosonic systems, the adaptive TDDMRG has been found to be a highly reliable real-time simulation method at economic computational cost, for example in the context of magnetization dynamics \cite{Gobert05}, of spin-charge separation \cite{Kollath05, Kleine08}, or far-from equilibrium dynamics of ultracold bosonic atoms \cite{Cramer08}. Recently, Zhao \textit{et al} have performed adaptive TDDMRG simulations for polaronic transport within a combined model of SSH and Hubbard model to take the on-site Coulomb interactions into account and found that that the velocity of the polaron is suppressed by the on-site Coulomb interaction $U$.\cite{Zhao08}

In this paper, combining the SSH and extended Hubbard model to take both the on-site Coulomb interactions and
nearest-neighbor electron-electron interactions into account, we simulate the motion of a polaron in
conjugated polymers under an applied external electric field by using
the adaptive TDDMRG for the $\pi$-electron part and
classical molecular dynamics (MD) for the lattice backbone part. The
aim of this paper is to give an exhaustive picture of polaron transport in conducting polymers at a theoretical level with all
relevant electron-electron interactions and correlations included
and to show how the on-site Coulomb interactions $U$ and
nearest-neighbor electron-electron interactions $V$ influence the
behavior of polaron transport in conducting polymers.

\section{Model and Methodology}
We use the well-known and widely used SSH Hamiltonian \cite{Su79,
Su80} combined with the extended Hubbard model (EHM), and include
the external electric field by an additional term:
\begin{equation}\label{H}
    H(t)=H_{el}+H_{E(t)}+H_{latt}
\end{equation}
This Hamiltonian is time-dependent, because the electric field $E(t)$ is explicitly time-dependent.

The $\pi$-electron part includes both the electron-phonon and the
electron-electron interactions,
\begin{equation}\label{H_el}
\begin{split}
    H_{el}=&-\sum_{n,\sigma}t_{n,n+1}(c_{n+1,\sigma}^{+}c_{n,\sigma}+h.c.)\\
& +\frac{U}{2}\sum_{n,\sigma}(c_{n,\sigma}^{+}c_{n,\sigma}-\frac{1}{2})(c_{n,-\sigma}^{+}c_{n,-\sigma}-\frac{1}{2})\\
&
+V\sum_{n,\sigma,\sigma'}(c_{n,\sigma}^{+}c_{n,\sigma}-\frac{1}{2})(c_{n+1,\sigma'}^{+}c_{n+1,\sigma'}-\frac{1}{2})
\end{split}
\end{equation}
where $t_{n,n+1}$ is the hopping integral between the $n$-th site
and the ($n+1$)-th site, while $U$ is the on-site Coulomb interaction
and $V$ denotes the nearest-neighbor electron-electron interaction.
Because the distortions of the lattice backbone are always within a
certain limited extent, one can adopt a linear relationship between
the hopping integral and the lattice displacements as
$t_{n,n+1}=t_0-\alpha(u_{n+1}-u_n)$ \cite{Su79, Su80}, where $t_0$
is the hopping integral for zero displacement, $u_n$ the lattice
displacement of the $n$th site, and $\alpha$ is the electron-phonon
coupling.

Because the atoms move much slower than the electrons, we treat the lattice backbone classically with the Hamiltonian
\begin{equation}\label{H_latt}
    H_{latt}=\frac{K}{2}\sum_{n}(u_{n+1}-u_n)^2+\frac{M}{2}\sum_{n}\dot{u}_n^2 \text{    },
\end{equation}
where $K$ is the elastic constant and $M$ is the mass of a site, such as that of a CH monomer for trans-polyacetylene.

The electric field $E(t)$ directed along the backbone chain is uniform over
the entire system. The field which is constant after a smooth
turn-on is chosen to be
  \begin{equation}\label{eq:t2}
    E(t)=\begin{cases}
    E_0\exp[-(t-T_C)^2/T_W^2],    &\text{for $t< T_C$, } \\
    E_0,    &\text{for $t\geq T_C$, }
    \end{cases}
  \end{equation}
where $T_C$, $T_W$  and $E_0$ are the center, width and strength of
the half Gaussian pulse. This field gives the following contribution to the
Hamiltonian:
\begin{equation}\label{H_Et}
    H_{E(t)}=\vert e\vert\sum_{n,\sigma}(na+u_n)(c_{n,\sigma}^{+}c_{n,\sigma}-\frac{1}{2})E(t)
\end{equation}
where $e$ is the electron charge and $a$ is the unit distance constant of the lattice.
The model parameters are those generally chosen for polyacetylene:
$t_0$=2.5 eV, $\alpha$=4.1 eV/\AA, $K$=21 eV/\AA$^2$, $M$=1349.14
eVfs$^2$/\AA$^2$, $a$=1.22 \AA.\cite{Su80} The results are expected
to be qualitatively valid for the other conjugated polymers. The
values of $T_C$ and $T_W$ ($T_C$=30 fs and $T_W$=25 fs) are taken
from the paper of Fu \textit{et al}.\cite{Fu06}

For the purpose of performing real-time simulation of both the evolution of quantum $\pi$-electron part and the classical movement of the chain backbone, we adopt a newly developed real-time simulation method in which classical molecular dynamics is combined with the adaptive TDDMRG. The main idea of this method is to evolve the $\pi$-electron part by the adaptive TDDMRG and move the backbone part by classical MD iteratively. Details about this method can be found in recent papers \cite{Zhao08, Ma08}.

\section{Results and discussion}
To investigate the dynamic properties of polarons, we simulate the
polaron transport process in a single model conjugated polymer chain under a uniform external
electric field. A polaron is different from a charged soliton in that it involves an unpaired electron and thus a nonzero $S=1/2$. In our calculations, we simulate a model chain containing $N=100$ monomers and $N_e=99$ $\pi$-electrons to present a polaron defect without degenerate ground states. The polaron is initially centered around site 30 and site 31 through imposing a constraint of reflection symmetry around the center between site 30 and site 31. Then, the dynamics of polaron transport in 160 femtoseconds (fs) is simulated by virtue of classical MD
combined with the adaptive TDDMRG.

\subsection{General picture of polaron transport}
\begin{figure}
\includegraphics[width =14 cm]{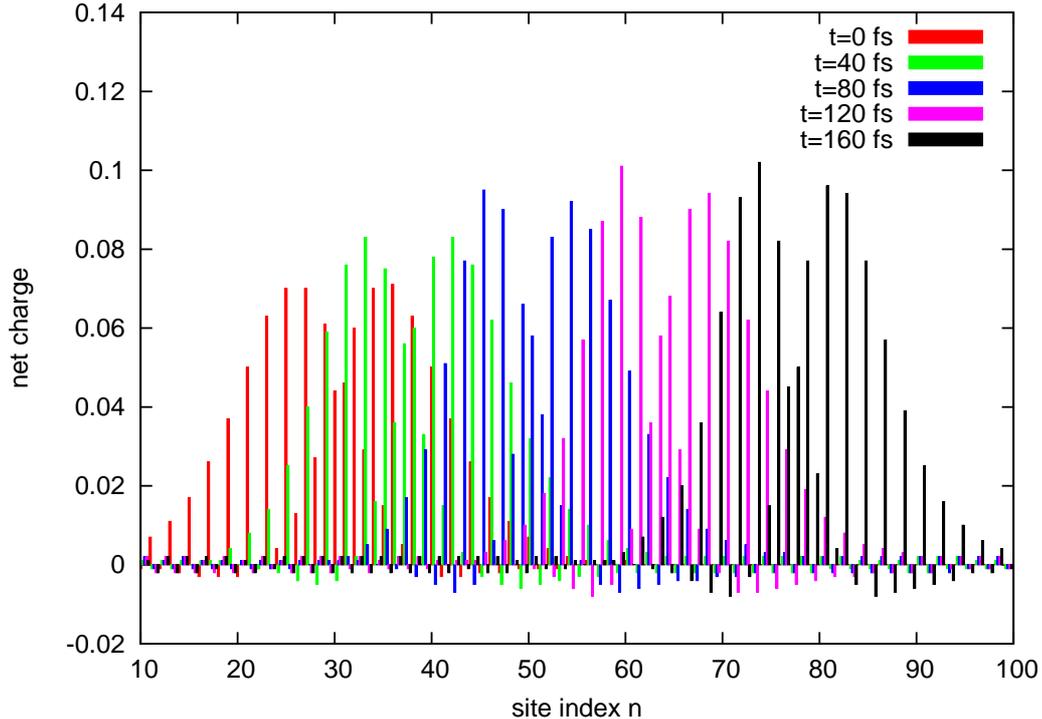}
\caption{\label{fig:c} Time evolution of net charge distribution in a polymer chain with the transport of a polaron under an external electric field. (The polaron moves from the left to the right.) ($E_0$=3.0 mV/$\text{\AA}$, U=2.0 eV, V=1.0 eV)}
\end{figure}
\begin{figure}
\includegraphics[width =14 cm]{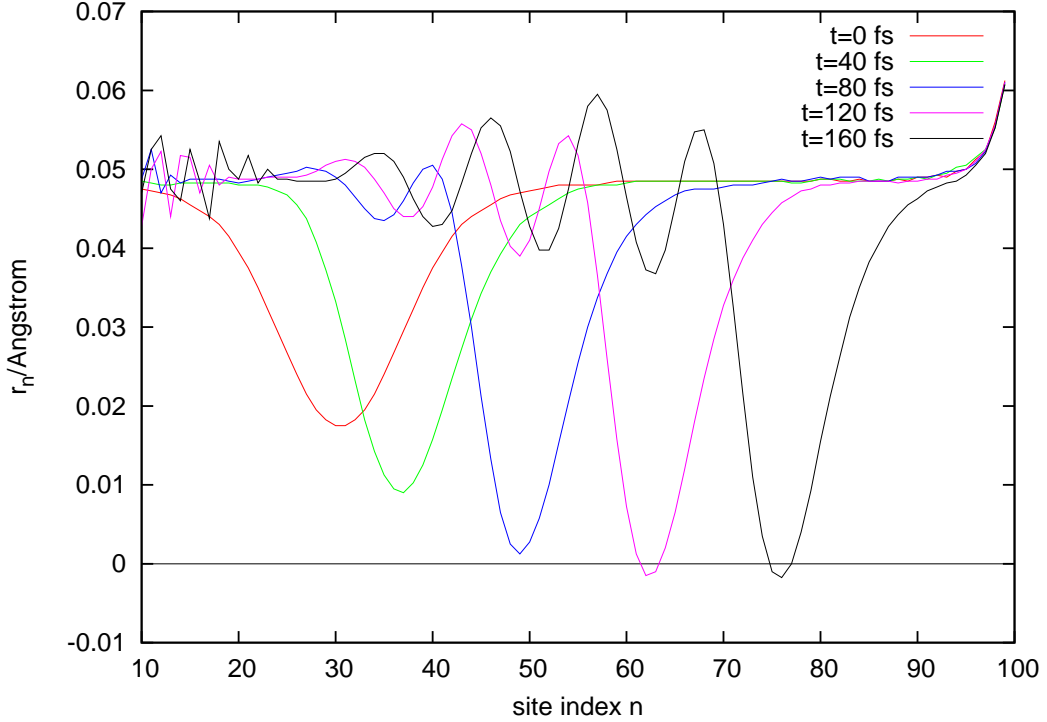}
\caption{\label{fig:g} Time evolution of the staggered bond order parameter $r_n$ in a polymer chain with the transport of a polaron under an external electric field. (The polaron moves from the left to the right.) ($E_0$=3.0 mV/$\text{\AA}$, U=2.0 eV, V=1.0 eV)}
\end{figure}
Firstly let us show the general time evolution
picture of polaron transport in a conducting polymer. The
time evolution of charge density and the staggered bond order
parameter $r_n=(-1)^n(2u_n-u_{n+1}-u_{n-1})/4$ in a polymer chain with a polaron defect are shown in Fig.~\ref{fig:c} and
Fig.~\ref{fig:g}. It can be clearly seen that both the charge density and the geometrical distortions don't localize at one monomer. On the contrary, the polaron defect spreads over a delocalized region with a length of tens of sites. One can also clearly see that during the entire time
evolution process the geometrical distortion curve and net charge
distribution shape for the polaron defect always stay coupled
and show no obvious dispersion. This implies that the polaron defect is an
inherent feature and the fundamental charge carrier in conducting
polymer. In Fig.~\ref{fig:g}, we can also see that a long lasting
oscillatory ``tail'' appears behind the polaron defect center. This ``tail'' is generated by the inertia of those monomers to fulfill energy and momentum conservation; they absorb the additional energy, preventing the further increase of the polaron velocity after a stationary value is reached. 

In order to evaluate the velocity of polaron transport, it is necessary to know the center postions of a polaron at different times. The center position of a defect can be derived from the charge density distribution picture or the bond length alternation pattern as
  \begin{equation}\label{eq:Po}
    x_{c,g}(t)=\begin{cases}
    N\theta_{c,g}(t)/2\pi,    &\text{if cos$\theta_{c,g}\geq0$ and sin$\theta_{c,g}\geq0$;} \\
    N(\pi+\theta_{c,g}(t))/2\pi,    &\text{if cos$\theta_{c,g}<0$;}\\
    N(2\pi+\theta_{c,g}(t))/2\pi,    &\text{otherwise, }
    \end{cases}
  \end{equation}
where $\theta_{c,g}$ is defined according to the charge density $\rho_n$ or the staggered bond order
parameter $r_n$ as
\begin{equation}\label{theta}
\begin{split}
    &\theta_c(t)=\text{arctan}\frac{\sum_n\rho_n(t)\text{sin}(2\pi n/N)}{\sum_n\rho_n(t)\text{cos}(2\pi n/N)},\\
& \theta_g(t)=\text{arctan}\frac{\sum_n(r_n(t)-r^0_n)\text{sin}(2\pi n/N)}{\sum_n(r_n(t)-r^0_n)\text{cos}(2\pi n/N)},
\end{split}
\end{equation}
in which $r^0$ is the dimerized lattice displacement of the pristine chain without defects. 

\begin{figure}
\includegraphics[width =8 cm]{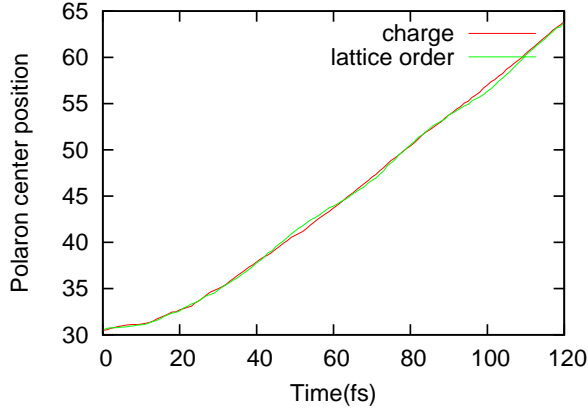}
\caption{\label{fig:Po_t} Temporal evolution of the center position of a polaron under an external electric field derived from both the charge density and the bond order parameter. ($E_0$=3.0 mV/$\text{\AA}$, U=2.0 eV, V=1.0 eV)}
\end{figure}
From Fig.~\ref{fig:c} and Fig.~\ref{fig:g}, one can also find that,
after the external electric field is smoothly turned on, the polaron is accelerated at
first and then moves with a nearly constant velocity as one entity
consisting of both the charge and the lattice deformation.
We calculate the postions of the center of a polaron during the transport process according to Eq.~\ref{eq:Po} and show the temporal evolution of the polaron center position in Fig.~\ref{fig:Po_t}. It can be clearly seen that the center position of the polaron increases nearly linearly with the increase of time after the initial short-time acceleration. This implies the polaron defect is transported with a nearly constant velocity and supports our observation from Fig.~\ref{fig:c} and Fig.~\ref{fig:g}. Therefore, only the stationary
velocity will be considered for polaron transport in the following discussions. Meanwhile, one also finds that the lattice distortion center always stays well together with the charge density center. But it should also be noticed that, under high electric field strength the deviation of the lattice distortion center position from the charge density center position may become much larger because the charge density moves much faster under high field and the lattice distortion cannot catch up with the charge density. This separation between the charge center and lattice distortion center may lead to the dissociation of a polaron. The dissociation of a polaron under high electric field strength will be discussed in the next section.

\subsection{Relationship between polaron transport velocity and electric field strength}
\begin{figure}
\includegraphics[width =8 cm]{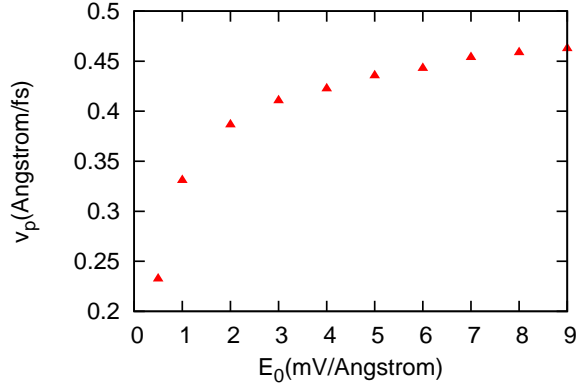}
\caption{\label{fig:E} The stationary velocity of a polaron $v_p$ as a function of the
electric field strength. (U=2.0 eV, V=1.0 eV) }
\end{figure}
The dependence of the stationary velocity of a polaron $v_p$ on the electric field
strength is shown in Fig.~\ref{fig:E}. We find that $v_p$ increases with increasing electric field strength. Apparently, the polaron velocity can easily exceed the sound speed ($V_S=\sqrt{4K/M}a/2$=0.15 \AA/fs) and the maximum velocity (about 3$V_S$) is reached at a high electric field strength ($\sim$ 9 mV/$\text{\AA}$). Moreover, a very interesting behavior of $v_p$ as a function of the
electric field strength is observed. An ohmic region where $v_p$ increases nearly linearly with
the electric field strength is found. This ohmic region extends approximately from 3 mV/$\text{\AA}$ to 9 mV/$\text{\AA}$ for the case of $U$=2.0 eV and $V$=1.0 eV. Below 3 mV/$\text{\AA}$ the polaron velocity increases nonlinearly and when $E_0\geq 10$ mV/$\text{\AA}$ the polaron becomes unstable and the charge will be decoupled from the lattice distortion.

\begin{figure}
\includegraphics[width =12 cm]{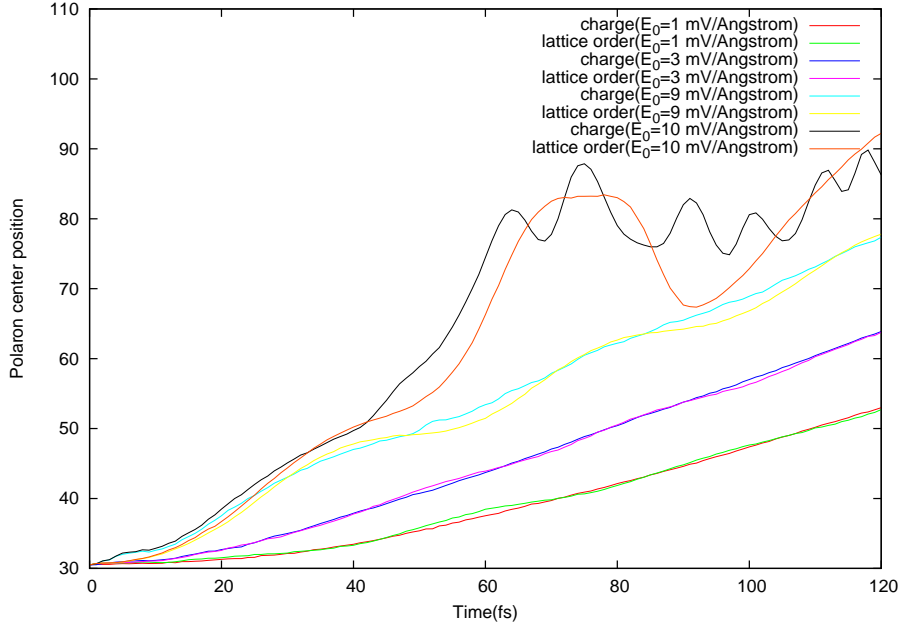}
\caption{\label{fig:Po_t-E} Temporal evolution of the center position of a polaron under different external electric field strengths. (U=2.0 eV, V=1.0 eV)}
\end{figure}
\begin{figure}
\includegraphics[width =14 cm]{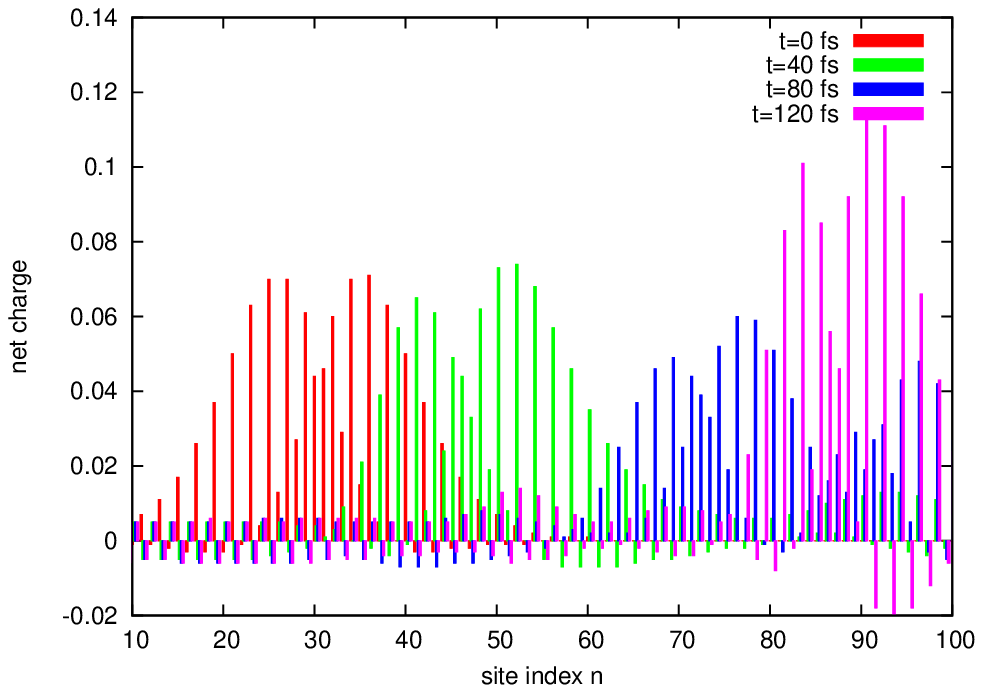}
\caption{\label{fig:c_E} Time evolution of net charge distribution in a polymer chain with the transport of a polaron under a high external electric field. (The polaron moves from the left to the right.) ($E_0$=10.0 mV/$\text{\AA}$, U=2.0 eV, V=1.0 eV)}
\end{figure}
\begin{figure}
\includegraphics[width =14 cm]{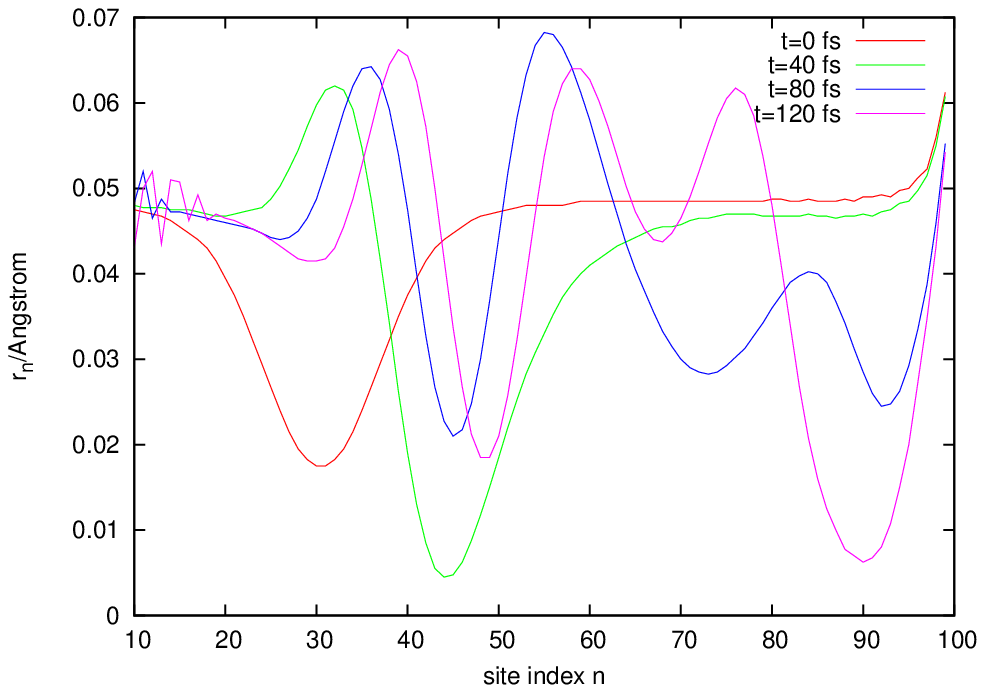}
\caption{\label{fig:g_E} Time evolution of the staggered bond order parameter $r_n$ in a polymer chain with the transport of a polaron under a high external electric field. (The polaron moves from the left to the right.) ($E_0$=10.0 mV/$\text{\AA}$, U=2.0 eV, V=1.0 eV)}
\end{figure}
We illustrate the temporal evolutions of the polaron center postion under different external electric field strengths. We can find that the lattice distortion center stays well together with the charge center under low electric field strengths. However, the deviation of the lattice distortion center from the charge density center becomes much larger under higher electric field strengths because the charge density moves much faster under high field and the lattice distortion cannot catch up with the charge density. Under the electric field of 10 mV/$\text{\AA}$, the distance between the two centers becomes even larger than several unit distances, implying that the polaron has been completely dissociated. This dissociation can be clearly seen from the
time evolution pictures of charge density and the staggered bond order
parameter in a polymer chain with a polaron defect under a external electric field of 10 mV/$\text{\AA}$ in Fig.~\ref{fig:c_E} and
Fig.~\ref{fig:g_E}. While the charge density accelerates to a high speed as shown in Fig.~\ref{fig:c_E}, the original lattice distortion stays far behind the charge density with a distance of more than tens of unit distances. The charge density will induce new lattice distortion in the nearby region around the charge density center, while the original lattice distortion can hardly move further, as illustrated in Fig.~\ref{fig:g_E}. Therefore, the polaron defect has become completely dissociated under the high electric field. Apparently, our calculated threshold field strength of around 10 mV/$\text{\AA}$ for polaron dissociation is in agreement with the value of 9.5 mV/$\text{\AA}$ obtained by SSH-EHM/UHF calculations by Di \textit{et al} \cite{Di07} and much closer to the experimental determined value of about 15 mV/$\text{\AA}$ \cite{Lin02} compared to the value of 6 mV/$\text{\AA}$ calculated by Rakhmanova and Conwell \cite{Rakhmanova99, Rakhmanova00} and 3.5 mV/$\text{\AA}$ calculated by Johansson and Stafstr\"{o}m \cite{Johansson01, Johansson04}. The large deviations of the calculated threshold field strength from the experimental value by the latter two groups are due to the fact that they considered only the SSH model and considered no electron-electron interactions in their studies, and the agreement of SSH-EHM calculations with experimental results shows again that the inclusion of electron-electron interactions is vital for the theoretical studies of charge carriers in conducting polymers. Another thing that should be noticed is that our $U$ value (2.0 eV) and $V$ value (1.0eV) are not the standard values for any special kind of polymers. Here we use these two values only for the purpose to show how a polaronic transport will generally evolve and whether the inclusion of electron-electron interactions is important for a reasonable theoretical simulation of the polaronic transport process. The detailed discussion of how the on-site repulsions $U$ and the nearest-neighbor interactions $V$ will influence the polaronic transport will be shown in the next sections.

\subsection{Influence of on-site repulsions $U$ on polaron transport velocity}
In order to study the influence of electron-electron interactions
on polaron transport, we focus on the stationary velocity of a polaron $v_p$
calculated with different electron-electron interaction strengths under the constraint that
the other parameters are fixed. As should be noticed that, the real conjugated polymer is with weak interactions. Strong interactions will lead to too strong charge polarizations which are unrealistic. Considering that we are only focusing on the
study of real conjugated polymer system, in this work we adopt only the
weak-interaction parameters ($U<2t, V<t$). 

\begin{figure}
\includegraphics[width =8cm]{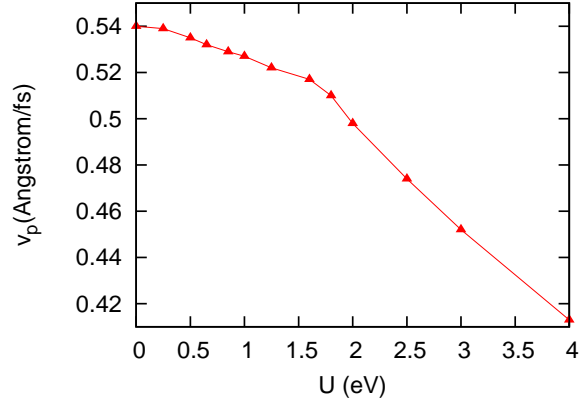}
\caption{\label{fig:v_U} The stationary transport velocity of a polaron as a function of different $U$ values. ($E_0$=3.0 mV/\AA, $V$=0.0 eV)}
\end{figure}
\begin{figure}
\includegraphics[width =12 cm]{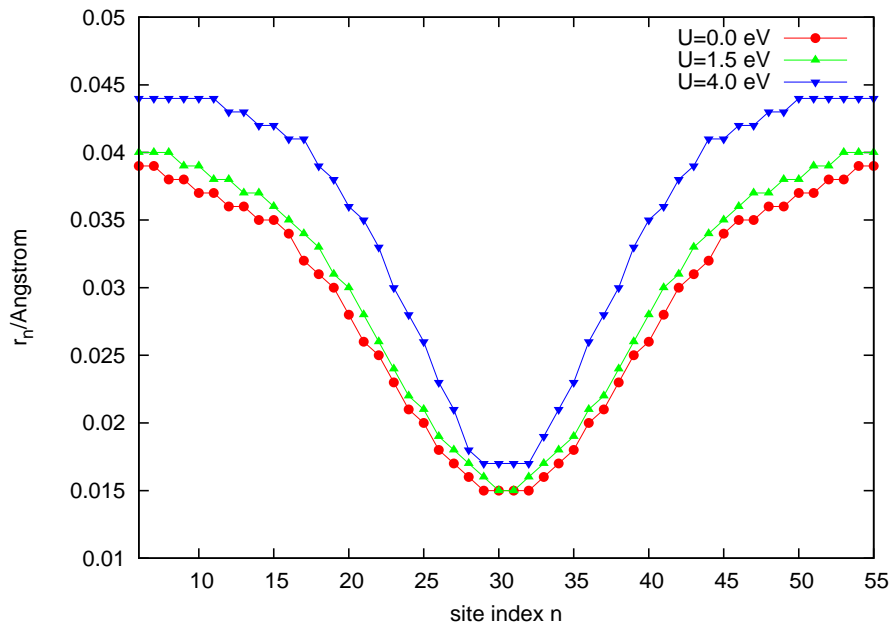}
\caption{\label{fig:g_U} The staggered bond order parameter $r_n$ of
a static polaron for several different $U$ values. ($V$=0.0 eV)}
\end{figure}
Firstly, we study the Hubbard model with only on-site Coulomb interactions $U$, \textit{i.e.}, $V=0$ in
the EHM. The dependence of the polaron stationary velocity $v_p$ on
the on-site Coulomb interactions $U$ is displayed in
Fig.~\ref{fig:v_U}. We find that $v_p$ decreases monotonically with the increasing $U$ value and its maximum value is achieved at $U$=0.0 eV. This can be easily understood because the variation of the charge carrier transport velocity is strongly related to the delocalization level of the charge carrier defect. The lattice tends to be occupied by one
electron per site when the on-site Coulomb repulsion $U$ increases. Therefore an increasing $U$ will lead to more localized charge density and a smaller defect width of the polaron and accordingly restrain the polaron transport. This analysis is supported by our calculated charge density width values of a static polaron $W_c$. $W_c$ is calculated according to the following formula:
\begin{equation}\label{Wc}
W_c=(\sum_n[n-X_c]^2\rho_n)^{1/2}.
\end{equation}
Through calculations we find the polaron width $W_c$ also decreases monotonically with an increasing $U$ value, from 10.10 ($U$=0.0 eV) to 9.68 ($U$=1.5 eV) and then to 7.62 ($U$=4.0 eV). This sequence is in agreement with that of the polaron stationary velocity $v_p$ and supports our analysis of the relationship between the polaron transport velocity and the delocalization level of a polaron. Actually, the change of the polaron delocalization can also be directly viewed through the geometrical picture of the static polaron defect. In Fig.~\ref{fig:g_U}, the staggered bond order
parameter $r_n$ of a static polaron calculated by different $U$ values is
shown. It can be clearly seen that the lattice dimerization is enhanced and the polaron width becomes narrower while the $U$ value increases. All these pictures show that the polaron transport is suppressed by the on-site Coulomb repulsions $U$. The monotonic decrease picture of polaron transport velocity with on-site Coulomb repulsions $U$ observed by us is in agreement with Di \textit{et al}'s UHF calculations \cite{Di07} and Zhao \textit{et al}'s recent adaptive TDDMRG results \cite{Zhao08}, but not in accordance to our recent studies on charged soliton transport, in which we found that charged soliton transport velocity is non-monotonic in $U$ \cite{Ma08}. The different behaviors of charge carrier transport with $U$ increasing is due to the different characteristics of charged solitons and polarons. It was found that the polaron defect is more delocalized than the charged soliton defect and the height of the charge density peaks in polarons is only roughly one half that of the charge density peaks in charged solitons.\cite{Ma06} In a more localized case with large charge densities, as in a charged soliton, small $U$ may favor the charge carrier transport because the increasing on-site repulsion for large charge densities can make the electron (or hole) to hop more easily to the next site. Therefore the different behaviors of charge carrier transport with $U$ increasing for charged solitons and polarons are reasonable.

\subsection{Influence of nearest-neighbor interactions $V$ on polaron transport velocity}
\begin{figure}
\includegraphics[width =8cm]{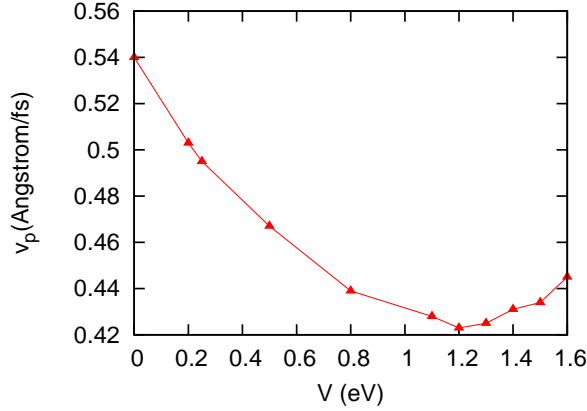}
\caption{\label{fig:v_V} The stationary transport velocity of a polaron as a function of different $V$ values. ($E_0$=3.0 mV/\AA, $U$=0.0 eV)}
\end{figure}
\begin{figure}
\includegraphics[width =9 cm]{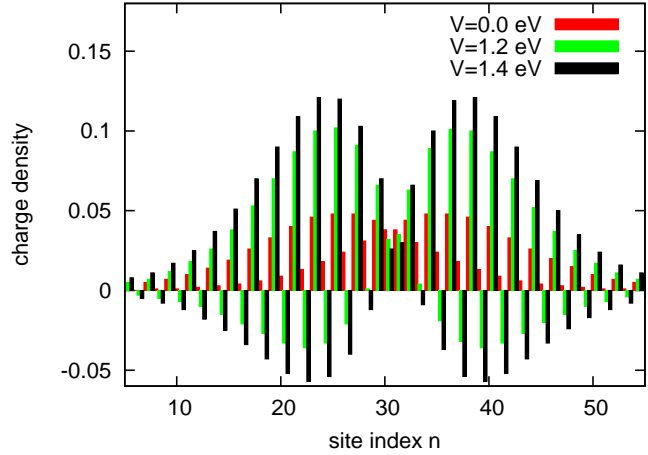}
\caption{\label{fig:c_V} The charge density of a static polaron as a function of the site
index calculated with different $V$ values. (U=0.0 eV)}
\end{figure}
\begin{figure}
\includegraphics[width =12 cm]{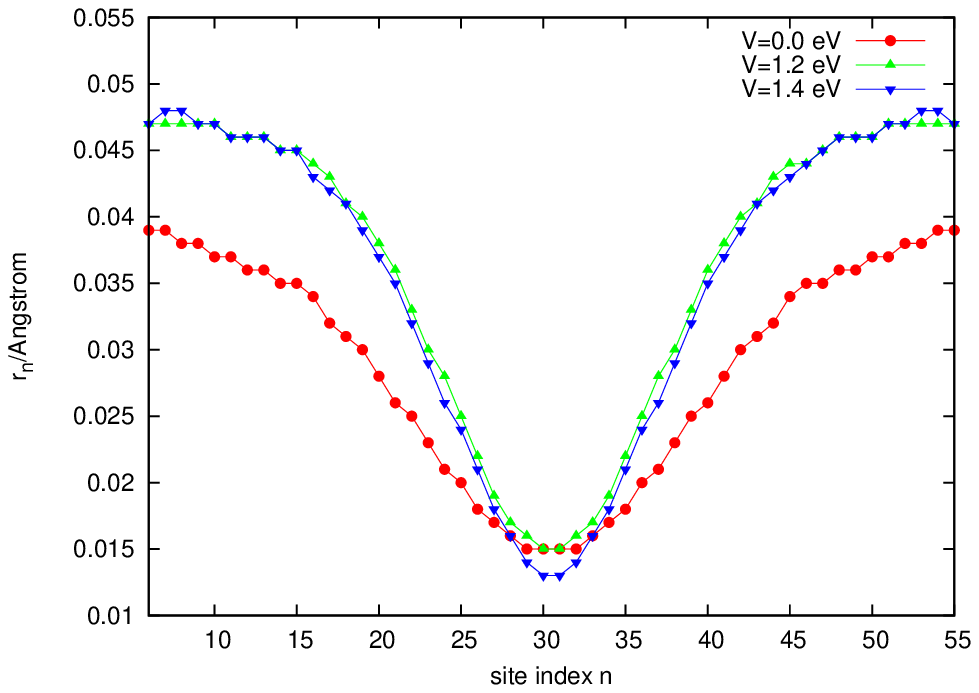}
\caption{\label{fig:g_V} The staggered bond order parameter $r_n$ of
a static polaron for several different $V$ values. ($U$=0.0 eV)}
\end{figure}
Secondly, we also study the influence of nearest neighbor
electron-electron interactions $V$ on the polaron transport process.
The dependence of the stationary velocity of a polaron $v_p$ on the $V$
values is displayed in Fig.~\ref{fig:v_V}, where $U$ values are
supposed to be frozen to be zero. This assumption is not realistic
because the on-site Coulomb interactions $U$ are normally much
stronger than the neighbor electron-electron interactions $V$; we make this assumption only for the purpose of
studying of the influence of $V$ on the polaronic transport process
without the effect of $U$. As can be seen in Fig.~\ref{fig:v_V}, the situation is completely different from the case of an on-site Coulomb interaction. Interestingly, $v_p$ is non-monotonic in $V$: it
decreases to a shallow minimum at $V\approx1.2$ eV gradually and then increases again. In fact, similar to the the case discussed above with varying $U$, the variation of $v_p$ with $V$ is also strongly related to changing the delocalization of the polaron defect. For small $V$, larger positive charge densities will be induced in the central part of a polaron defect while the induced negative charge densities at nearest neighbor sites are still relatively small, as shown in Fig.~\ref{fig:c_V}. Therefore the polaron defect tends to be more localized and consequently the polaron transport becomes more difficult. So, $v_p$ decreases while the value of $V$ increases from 0.0 eV to 1.2 eV as displayed in Fig.~\ref{fig:v_V}. However, when nearest neighbor electron-electron interactions $V$ increase further and
become dominant, very large charge polarization will be induced. Therefore, the electron-hole attraction between opposite charge densities at nearest-neighbor sites will contribute much more significantly to the polaron system and favor the hoppings of the accumulated electrons (or holes) in the central part of a polaron defect to the neighbor sites. As shown in in Fig.~\ref{fig:c_V}, the two positive charge density peaks are separated further when $V$ increases from 1.2 eV to 1.4 eV. This change leads to a more delocalized polaron defect and accordingly the increase
of $v_p$. Our calculated polaron charge density width results, namely that $W_c$ decreases gradually from 10.10 ($V$=0.0 eV) to 9.37 ($V$=1.2 eV) and then increase gradually to 9.89 ($V$=1.4 eV), support our analysis well. In order to directly view the change of delocalization level of the polaron defect through geometrical pictures,  we also show the staggered bond order parameter $r_n$ of a static polaron calculated with different $V$ values in Fig.~\ref{fig:g_V}. It is clearly shown that the lattice dimerization is enhanced and the polaron width becomes narrower with $V$ for small $V$. We also find that the polaron becomes more delocalized with a smaller $r_n$ minimum while $V$ increases further from 1.2 eV to 1.4 eV. The change of polaron delocalization illustrated from the geometrical picture is in accordance to that has been shown in the charge density picture, verifying that small $V$ suppresses the polaronic transport while large $V$ favors the polaron transport. This non-monotonic picture of polaron transport velocity in $V$ is similar to that has been observed in charged soliton transport \cite{Ma08}.

\subsection{Influence of both the on-site Coulomb interactions $U$ and the nearest-neighbor interactions $V$ on polaron transport velocity}
\begin{figure}
\includegraphics[width =12 cm]{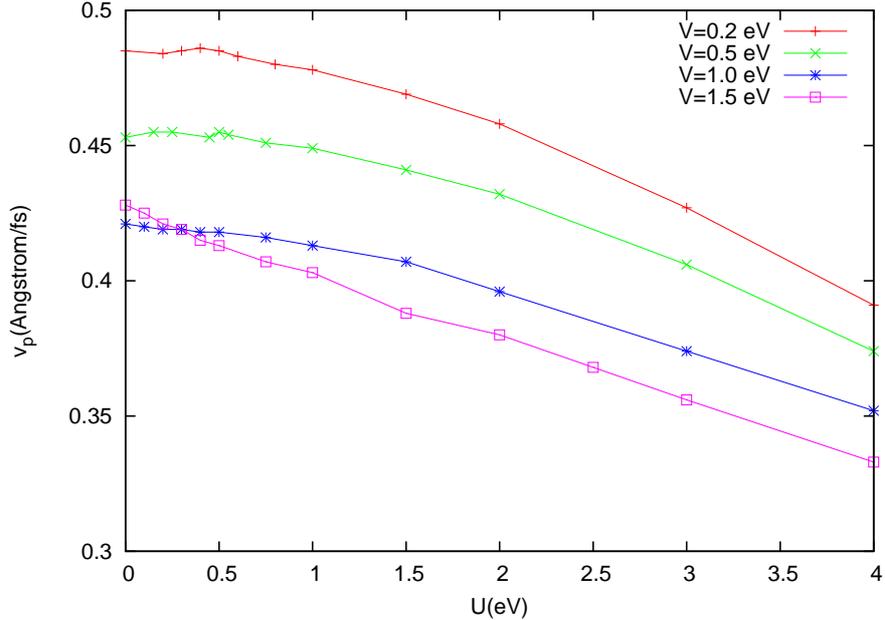}
\caption{\label{fig:v_f} The stationary transport velocity of a polaron as a function of $U$
for fixed values of $V$. ($E_0=3.0$ mV/\AA)}
\end{figure}
Furthermore, we consider the realistic case of conducting polymers in which both the on-site Coulomb
interactions $U$ and the nearest-neighbor interactions $V$ are taken into account.
Fig.~\ref{fig:v_f} shows the stationary transport velocity of a polaron $v_p$ as a
function of $U$ for fixed values of $V=0.2 $ eV, $V=0.5 $ eV, $V$=1.0 eV and
$V=1.5$ eV. Generally, the $v_p$ behavior with variated $U$ and fixed $V$ values is quite similar to that has been illustrated in Fig.~\ref{fig:v_U} in which $V$ is frozen to be 0.0 eV. It can be found that $v_p$ decreases monotonically with $U$. Meanwhile, we can also find that, $v_p$ seems to decrease more rapidly in large $V$ cases than in small $V$ cases for small $U$. Larger nearest-neighbor electron-electron interactions $V$ will lead to larger charge polarizations and of course the increasing on-site Coulomb repulsions $U$ will suppress the transport of a polaron associated with larger charge polarizations more significantly comparing to the case with smaller charge polarizations. Therefore the suppression of polaron transport by the on-site Coulomb repulsions $U$ is more remarkable in large $V$ cases than in small $V$ cases. The monotonic decrease picture of polaron transport velocity with on-site Coulomb repulsions $U$ observed by us is opposed to the non-monotonic picture obtained by Di \textit{et al} through UHF calculations \cite{Di07}. They found a local extremum of the stationary velocity of the polaron can be achieved at $U\approx 2V$. The difference between the results of the adaptive TDDMRG and UHF calculations is due to the fact that the latter method ignores the important electron correlation effect. Previous theoretical static studies have found that neglecting electron correlation effects will lead to an underestimation of the defect width as well as an overestimation of charge polarization.\cite{Fonseca01, Oliveira03, Champagne04, Monev05} In a more localized case with large charge densities, presented by UHF calculations without the inclusion of electron correlation effect, introducing a small $U$ may favor the charge carrier transport because the increasing on-site repulsion for large charge densities can make the electron (or hole) to hop more easily to the next site. Therefore, the difference between TDDMRG calculations and UHF calculations shows again that the electron correlation effect plays an important role in describing the charge carrier properties.

\section{Summary and conclusion}
For a model conjugated polymer chain initially holding a polaron defect, described using the combined SSH-EHM model extended with an additional part for the influence of an external electric field, we 
have studied the dynamics of polaron transport through this chain by virtue of
simulating the backbone monomer
displacements with classical MD and evolving
the wavefunction for the $\pi$-electrons with the adaptive TDDMRG.

It is found that after a smooth turn-on of the external electric field
the polaron is accelerated at first up to a stationary constant velocity as one entity consisting of
both the charge and the lattice deformation. During the entire time
evolution process the geometrical distortion curve and charge
distribution shape for the polaron defect always stay well coupled
and show no dispersion under low external electric fields, implying that the polaron defect is an
inherent feature and the fundamental charge carrier in
conducting polymer. The dependence of the stationary velocity of a polaron $v_p$ on the external electric field
strength is also studied, and an ohmic region where $v_p$ increases linearly with
the field strength is found. Values beyond the speed of sound are achievable. This ohmic region extends approximately from 3 mV/$\text{\AA}$ to 9 mV/$\text{\AA}$ for the case of $U$=2.0 eV and $V$=1.0 eV. Below 3 mV/$\text{\AA}$ the polaron velocity increases nonlinearly and in high external electric field strengths the polaron will become unstable and dissociate. Our calculated threshold field strength for polaron dissociation is around 10 mV/$\text{\AA}$ for the case of $U$=2.0 eV and $V$=1.0 eV. It is in good agreement with experimental results and more physically intuitive than previous SSH calculations which take no electron-electron interactions into account.

The influence of electron-electron interactions (both the on-site Coulomb
interactions $U$ and the nearest-neighbor interactions $V$) on polaron transport are
investigated in detail. In general, the increase of the on-site Coulomb
interactions $U$ makes the lattice tend to be occupied by one electron per site and accordingly suppress the polaron transport. Therefore,  $v_p$ decreases monotonically with $U$. Meanwhile, small $V$ values are not beneficial to the polaronic motion because they induce a more localized defect distribution, and due to the induced large charge polarization accompanied with large nearest-neighbor attractions large $V$ values favor the polaron transport. When $U$ and $V$ are considered at the same time, $v_p$ also decreases monotonically with the increasing $U$ value and $v_p$ decreases more rapidly for small $U$ in large $V$ cases than in small $V$ cases.

\section*{Acknowledgment}
HM is grateful to Andrej Gendiar for helpful
discussions. HM also acknowledges the support by an Alexander von
Humboldt Research Fellowship.

\end{document}